\documentclass[11pt]{article}
\usepackage[dvips]{graphicx}

\begin{document}
\title{\Large \bf Neutralino inelastic scattering with 
subsequent detection of nuclear $\gamma$ rays}
\author{ \\ J.\ Engel$^1$ and P. Vogel$^2$ \\ 
{\small $^1$Department of Physics and Astronomy, CB3255,
University of North Carolina,} \\ {\small
Chapel Hill, North Carolina 27599} \\
{\small $^2$Physics Department 161-33, Caltech, Pasadena, CA 91125}}

\maketitle \begin{abstract}

{\normalsize We consider the potential benefits of searching for 
supersymmetric dark-matter
through its inelastic excitation, via the ``scalar current", of low-lying 
collective nuclear states in a detector.  If such states live long enough so 
that the $\gamma$
radiation from their decay can be separated from the signal due to nuclear
recoil, then background can be dramatically reduced.  We show how the
kinematics of neutralino-nucleus scattering is modified when the nucleus
is excited and derive expressions for the form factors associated with
exciting collective states.  We apply these results to two specific cases:  1)
the $I^{\pi} = 5/2^+$ state at 13 keV in $^{73}$Ge, and 2) the rotational and
hence very collective state $I^{\pi} = 3/2^+$ at 8 keV in $^{169}$Tm (even
though observing the transition down from that state will be difficult).  In
both cases we compare the form factors for inelastic scattering with those for
elastic scattering.  The inelastic cross section is considerably
smaller than its elastic counterpart, though perhaps not always prohibitively
so.}

\end{abstract} 

A number of groups are trying to detect weakly interacting dark matter, one of
the most promising candidates for which is the supersymmetric ``lightest
neutralino".  A popular approach is to try to observe the scattering of
these particles on nuclear targets in low-background laboratory experiments.
The signature of neutralino-nucleus scattering is the low-energy recoil of the
nucleus in a detector.  Since the scattering rate is expected to be tiny,
background is the main factor limiting sensitivity, even when low itself.

Supersymmetric dark matter is reviewed in ref.\ \cite{Griest}.  Here we are
interested only in the nuclear physics aspects of this problem, and in
particular in the possibility of detecting inelastic scattering, thereby
dramatically reducing background.  (The nuclear physics of dark matter
detection is reviewed in ref.\ \cite{us}.)  The work was inspired by
questions from researchers in the field\cite{Gaitskell,Avignone}.

Though inelastic scattering of neutralinos has been considered before, notably
in ref.\ \cite{Ellis}, the focus was on spin-dependent scattering.  The
authors discussed low-lying excited states in stable nuclei with large
measured M1 matrix elements; later, ref.\ \cite{Ejiri} reported an upper limit
of $9.8 \times 10^{-2}$ counts/kg/day (at 90\% CL) for
the inelastic excitation of the $7/2^+$ state at 57.6 keV in $^{127}$I.  It
has since become clear, however\cite{scalar}, that spin-independent scattering
will almost always occur with greater probability than its spin-dependent
counterpart.  We therefore focus here
on the possibility of excitation by the scalar current, where the relevant
multipole is E2 instead of M1.  Collective E2 transitions, of which there are 
many, may allow the scalar current to be even more effective.

Of course there is a price to pay for the extra $\gamma$-ray in the signal
from inelastic scattering:  the cross section is noticeably smaller than the
elastic one.  As we explain below, this is caused here not so much by the
kinematics discussed in ref.\ \cite{Ellis} --- E2 excitations can often be
found lower in the spectrum than M1 excitations --- or by the factor $qR$ that
enters higher multipoles, but rather by a considerable reduction in 
``coherence"
from elastic scattering, even when collective nuclear states are excited.
Collective excitations of the nucleus generally involve valence nucleons, of
which there are more than the (effectively) one that participates in
spin-dependent scattering, but still far fewer than the $A$ that are involved
in elastic scattering.  Thus, though we gain in some ways by considering the
scalar current, we will still not obtain cross sections that approach those
from elastic scattering.  We quantify this remark below.

Let us consider kinematics first.  A particle of mass $M_X$ 
moves with velocity $v$ and scatters on a stationary target of mass $M_A$. 
After the scattering the target has $E_{\rm exc}$ of excitation energy, i.e.\ 
its mass is $M_f = M_A + E_{\rm exc}$.  The momentum transfer is
\begin{equation}
\vec{q}^{~2} = M_X^2 | \vec{v} - \vec{v}' |^2 = 
M_X^2 [v^2 + v'^2 -2vv'\cos(\theta)] ~,
\end{equation}
where $\theta$ is the scattering angle and $v'$ the final velocity of the 
scattered particle. The energy transfer is
\begin{equation}
\omega = M_X (v^2-v'^2)/2 = E_{\rm recoil} + E_{\rm exc} = 
\frac{\vec{q}^{~2}}{2M_f }+ E_{\rm exc} ~.
\end{equation}
The minimum and maximum momentum transfer, and thus also the minimum and 
maximum recoil energy $E_{\rm recoil} = q^2/2M_f$, correspond to $\cos(\theta) 
= \pm 1$.  Eliminating $v'$ we obtain a quadratic equation for $q^2$ which 
gives
\begin{equation}
q_{\frac{\rm max}{\rm min}} = \mu v 
\left( 1 \pm \sqrt{1 - \frac{2 E_{\rm exc}}{\mu v^2}} \right) ~,
\label{eq:qmin}
\end{equation}
where $\mu = M_X M_f / (M_X+ M_f)$ (we can neglect the small difference 
between $M_A$ and $M_f$ here) is the reduced mass. Thus, for the inelastic 
process to occur at all, we must have $E_{\rm exc} < \mu v^2/2$. (Note that 
$\mu v^2/2$ is less than the neutralino kinetic energy, since $\mu < M_X$.)
To obtain the scattering rate of neutralinos with some velocity distribution
at a fixed momentum transfer $q$ (or recoil energy $E_{\rm recoil}$), we have 
to integrate over the velocity distribution from minimum velocity
\begin{equation}
v_{\rm min} = \frac{q}{2 \mu} + \frac{E_{\rm exc}}{q} ~.
\end{equation}
At the same time, for inelastic scattering there is an absolute minimum of 
momentum transfer, $q = \sqrt{2 \mu E_{\rm exc}}$.
 
Turning to the nuclear matrix elements that govern the cross section, we have 
from eqs.\ (4.24), (4.25) of ref.\ \cite{us} (generalized to transitions from 
$J \rightarrow J' \neq J$)
\begin{equation}
 \frac{{\rm d}\sigma}{{\rm d} q^2}  =  \frac{ 8 G_F^2 }{(2J+1) v^2 } S_S(q) ~,
\end{equation}
where the form factor for initial and final states of the same 
parity\footnote{One might imagine $E1$-like transitions between low-lying 
states 
of opposite parity, but for nuclear-structure reasons their strengths are 
notoriously small.} is
\begin{equation}
S_S(q)  =   \sum_{L~{\rm even}} | \langle J' || {\cal C}_L(q) || J \rangle |^2 
~,
\end{equation}
and
\begin{equation}
\label{eq:sph}
{\cal C}_{LM} (q)  =  \sum_i c_0 j_L (qr_i) Y_{L,M}(\hat{r}_i) ~.
\end{equation}
The summation over $L$ is restricted by 
$|J - J'| \le L \le J + J'$ and
the lowest allowed $L$ generally contributes most. For appropriate values of 
$J$ and $J'$ this value will correspond to the $L=2$ quadrupole mode, which 
also has 
the advantage of producing collective excitations of the nuclear surface; we 
denote the 
associated form factor by $S_2 (q)$.
We have lumped into the constant $c_0$ all the particle physics aspects of the
problem except the overall scaling $G_F^2$. In the ratio of inelastic to 
elastic form factors the constant $c_0$ drops out.

To calculate the matrix elements in eq.\ (\ref{eq:sph}) we have to know 
something about the structure of the initial and final states. 
The $q \rightarrow 0$ limit 
of the matrix element in eq.\ (\ref{eq:sph}) for $L=2$ can be measured in the 
Coulomb excitation or electromagnetic decay of the excited state.  The rates 
of these processes are usually expressed in terms of the quantity
\begin{equation}
B(E2, J \rightarrow J' ) = | \langle J' || e r^2 Y_2 || J \rangle |^2/ (2J + 
1) ~.
\end{equation}
Let us first consider the attractive $9/2^+ \rightarrow 5/2^+$ 
excitation in $^{73}$Ge.  That isomeric excited state at 13 keV has a long 
half-life (2.95$\mu$s) and a rather large $B(E2)$ (23 
Weisskopf units for the $\gamma$-decay transition $5/2^+ \rightarrow 9/2^+$).  
We make one crude but reasonable assumption here: that the transition density 
for the excitation is concentrated at the nuclear surface, as if the excited 
state were a vibration.  Then we have
\begin{equation}
B(E2, J \rightarrow J' ) \simeq e^2 \rho_0^2 R^4 \langle A_{\rm ang} \rangle^2 
/ (2 J + 1) ~.
\end{equation}
where $R$ is the nuclear radius, $\rho_0$ the proton density, and $A_{\rm 
ang}$ the matrix element of the angular factors. 

With the same assumptions we can write the form factor $S_2 (q)$ for the 
inelastic neutralino $J \rightarrow J'$  scattering as
\begin{equation}
S_2(q) \approx c_0^2 | \langle J' || j_2(qr) Y_2 || J \rangle |^2 = 
c_0^2 \left( \frac{A}{Z} \rho_0 \right )^2 j_2(qR)^2  \langle A_{\rm ang} 
\rangle^2 ~,
\end{equation}
where the factor $A/Z$ comes from the additional assumption that the neutron 
and proton densities are proportional.  Using the known $B(E2)$ we can 
rewrite the above as
\begin{equation}
\label{eq:s2}
S_2(q) = c_0^2 \frac{A^2}{Z^2} (2 J + 1) j_2(qR)^2 \frac{B(E2)}{e^2 R^4} ~.
\label{eq:vibr}
\end{equation}  The $S_2$ form factor can then be compared to the form factor
for elastic scattering, which is governed by the operator 
${\cal C}_{00} \equiv c_0 \sum_i j_0(qr_i) Y_{00}(\hat{r}_i)$. 
A constant density inside the nuclear radius and the relation
\begin{equation}
\int_0^R j_0(qr) r^2 dr = \frac{R^2}{q}j_1(qR)
\end{equation}
give
\begin{equation}
\label{eq:sel}
S_{el}(q) = c_0^2  (2J + 1)^2 A^2 \frac{9 j_1(qR)^2}{4 \pi (qR)^2} ~.
\end{equation}
The ratio of inelastic to elastic cross sections, 
$S_2(q)/S_{el}(q)$, from eqs.\ (\ref{eq:s2}) and (\ref{eq:sel}), is 
independent of the constant $c_0$.

Figure 1 (the upper panel) shows the elastic and inelastic form factors as a
function of the recoil energy $E_{\rm recoil}$, normalized to the elastic form
factor at $q=0$ (i.e.\ $E_{\rm recoil}=0$).  The inelastic form factor in fact
begins at a finite $E_{\rm recoil}$ related to the minimum momentum transfer
in eq.\ (\ref{eq:qmin}).  The largest $E_{\rm recoil}$ we consider, 140 keV,
corresponds to neutralinos of mass $\approx$ 60 GeV, (the mass indicated by a
recent experiment \cite{Bernabei}) moving with the galactic escape velocity,
650 km/s.  For inelastic scattering, eq.\ (\ref{eq:qmin}) restricts the
$E_{\rm recoil}$ to less than about 127 keV.  At low recoil energies, $E_{\rm
recoil} \le 30$ keV, the inelastic form factor is small because the spherical
Bessel function $j_2(x)$ is proportional to $x^2/15$ for small $x$.  Even at
larger recoil energies, however, the inelastic form factor is down from the
elastic one by a factor of 100 - 1000.  Only near the zero of the function
$j_1(x)$, which corresponds to $E_{\rm recoil} \sim 220$ keV in Ge, is the
inelastic cross section larger than the elastic one.  The small inelastic
cross section is caused by the absence of the coherence factor $A^2$ (which
appears divided by $Z$ in eq.\ (\ref{eq:s2}) only to renormalize 
the density).  The
collectivity of the E2 transition, which as noted above is restricted to the
nuclear surface, cannot fully compensate this loss.  Thus, while the sharp
$\gamma$ ray in the signal is undeniably beneficial, the expected count rate
is substantially smaller than in elastic scattering.  To further quantify
this statement we evaluate the total elastic and inelastic cross sections for
neutralinos with $M_X$ = 60 GeV and a Maxwellian velocity distribution
($\bar{v}$ = 220 km/s) terminated at the galactic escape velocity (650 km/s).
The result for an ideal detector is
\begin{equation}
\frac {\langle \sigma^{\rm inelastic} \rangle}{\langle \sigma^{\rm elastic} 
\rangle} = 2.8 \times 10^{-5} ~.
\end{equation}
A real detector will have some threshold in recoil energy below which it
is not sensitive.  The elastic form factor is largest at low recoil while
the inelastic form factor is completely negligible there; excluding events 
with energies below the lower limit will therefore increase the ratio above.
In a detector with a 10 keV threshold, the ratio is 
\begin{equation} 
\frac {\langle \sigma^{\rm inelastic} \rangle}{\langle \sigma^{\rm elastic} 
\rangle}_{\rm from~10~keV} = 5.7 \times 10^{-5}~,
\end{equation}
still a rather small number.

\begin{figure}[hbt]
\begin{center}
\includegraphics[width=10cm]{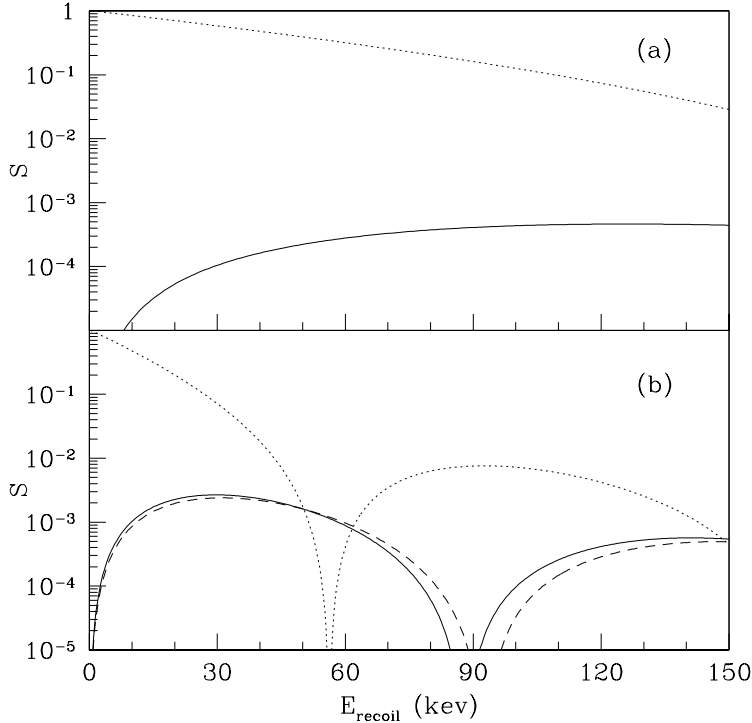}
\end{center}
\caption[Figure 1]{The quantities $S = \sigma (q)/\sigma_{\rm elastic} (q=0)$ 
for elastic (dotted lines) and inelastic (full lines) neutralino
scattering. The upper panel (a) is for $^{73}$Ge and the lower panel (b) for 
$^{169}$Tm.  The dashed line in (b) is the inelastic $S$ evaluated with
eq.\ (\ref{eq:s2}), which is less accurate than eq.\ (\ref{eq:s2r}).}
\end{figure}

Are there circumstances in which the reduction is not so dramatic and an
experiment more desirable?  For this to be the case, there must exist a
low-lying (not much more than 20 keV) excited state with a very collective E2
transition.  This state must live sufficiently long so that its deexcitation
can be separated in time from the signal caused by the recoil kinetic energy.
Finally, to eliminate the need for isotope enrichment, the target nucleus 
should be the only stable isotope of the 
element it represents. 

A quick search of the Table of Isotopes\cite{TI} reveal that these
conditions are not so easy to fulfill.  In fact, we found only one nucleus,
$^{169}$Tm, that comes close.  Its rotational $3/2^+$ state at 8.4 keV has a
half-life of 4.1 ns and very collective $B(E2; 3/2^+ \rightarrow
1/2_{g.s.}^+)$ of 226 Weisskopf units.  Detecting inelastic scattering
to this state will be difficult; its excitation energy is too
low and its half-life too short.  Nevertheless, we evaluated the corresponding
form factor to see what kind of count rates we could expect. 

In nuclei with permanent deformation the $B(E2)$ values are related
to the expectation value of $r^2Y_{20}$ in the intrinsic frame of the 
nucleus, 
which in turn follows from the deformation parameter $\beta$:
\begin{equation}
\langle r^2Y_{20}^{\rm intr}\rangle = \frac{3ZeR_0^2}{4\pi} \beta \left( 1 
+\frac{2}{7} \sqrt{\frac{5} {\pi}} \beta + \ldots \right)~.
\end{equation}
We can also write the intrinsic-frame expectation value of the operator
${\cal C}_{20}$ in eq.\ (\ref{eq:sph}) in terms of $\beta$: 
\begin{equation}
\langle {\cal C}_{20}^{\rm intr} (q) \rangle = 
\frac{3A c_0}{4\pi} \beta \left( j_2(qR_0) + \frac{1}{14}
\sqrt{\frac{5}{\pi}} \left[ qR_0 j_1(qR_0) - j_2(qR_0) \right] \beta 
+\ldots \right)~.
\label{eq:rot}
\end{equation}
To relate the inelastic form factor to 
the $B(E2)$ value, we use the expressions for rotational states:
\begin{equation}
B(E2; J,K \rightarrow J',K) =  \langle r^2Y_{20}^{\rm intr}\rangle^2 
\langle JK20|J'K \rangle^2 ~,
\end{equation}
and
\begin{equation}
\label{eq:s2r}
S_2(q; J,K \rightarrow J',K) = 
\langle C_{20}^{\rm intr} (q)\rangle ^2 (2J + 1) \langle JK20|J'K \rangle^2 ~.
\end{equation}
(The quantum number $K$, the angular-momentum projection on the nuclear 
symmetry axis, is $1/2$ for $^{169}$Tm.)
To leading order in $\beta$, these relations give the same result as eq.\ 
(\ref{eq:s2}).  The terms of order $\beta^2$ supply about a 10\% correction in 
$^{169}$Tm, which has $\beta \approx 0.3$.

The form factors for $^{169}$Tm appear in the lower panel of Fig.\ 1.  The
maximum recoil energy for a 60 GeV neutralino with the galactic-escape
velocity is now 112 keV.  The inelastic form factor, as expected, is not as
suppressed compared to its elastic counterpart as in $^{73}$Ge; the factor is
less than 100 in the broad maximum of the inelastic form factor at $\approx$
30 keV recoil energy.  The ratio of the total cross sections integrated from
the lowest possible momentum transfer is now \begin{equation} \frac {\langle
\sigma^{\rm inelastic} \rangle}{\langle \sigma^{\rm elastic} \rangle} = 1.5
\times 10^{-3}~, \end{equation} and increases to \begin{equation} \frac
{\langle \sigma^{\rm inelastic} \rangle}{\langle \sigma^{\rm elastic}
\rangle}_{\rm from~10~keV} = 5.9 \times 10^{-3} \end{equation} when integrated
from a 10-keV threshold.  To relate these results to those in Ge, one must
recall that the normalizing factor, $S_{el} (q=0)$, scales like $A^2$, i.e.\
it is larger for $^{169}$Tm than for $^{73}$Ge by $(169/73)^2$.  With a 10 keV
threshold, the integrated inelastic cross section per kg of material in
$^{169}$Tm is therefore suppressed with respect to the elastic cross section
in $^{73}$Ge by less than 100, a number that may not be so intimidating.

In conclusion, we have examined neutralino inelastic scattering to collective
states with large $B(E2)$ values.  We have shown how to evaluate the form
factors and presented examples.  While the search for inelastic
neutralino scattering offers an opportunity to suppress most background, it
also leads to a considerable reduction of the expected signal.


\end{document}